\documentclass[aps,prb,10pt,twocolumn]{revtex4-1}
\usepackage{bm}
\usepackage{amsmath}
\usepackage{amsfonts}
\usepackage{amssymb}
\usepackage{graphicx}
\usepackage[normalem]{ulem}

\usepackage{color}

\definecolor{Amber}{rgb}{1.0, 0.49, 0.0}

\begin{document}

\title{Janus Monolayers of Magnetic Transition Metal Dichalcogenides as \\
an {\it All-in-One} Platform for Spin-Orbit Torque}

\author{Idris Smaili$^1$}
\email{idris.smaili@kaust.edu.sa}
\author{Slimane Laref$^2$}
\email{slimane.laref@kaust.edu.sa}
\author{Jose H. Garcia$^{3}$ }
\author{Udo Schwingenschlogl$^{2}$}
\author{Stephan Roche$^{3,4}$ }
\author{Aur\'elien Manchon$^{1,2,5}$}
\email{aurelien.manchon@kaust.edu.sa}
\affiliation{$^1$King Abdullah University of Science and Technology (KAUST), Computer, Electrical, and Mathematical Science and Engineering Division (CEMSE), Thuwal 23955-6900, Saudi Arabia\\ $^2$King Abdullah University of Science and Technology (KAUST), Physical Science and Engineering Division (PSE), Thuwal 23955-6900, Saudi Arabia\\
$^{3}$Catalan Institute of Nanoscience and Nanotechnology (ICN2), CSIC and BIST,
Campus UAB, Bellaterra, 08193 Barcelona, Spain\\
$^4$ICREA--Instituci\'o Catalana de Recerca i Estudis Avan\c{c}ats, 08010 Barcelona, Spain\\
$^5$ Aix-Marseille Univ, CNRS, CINaM, Marseille, France.}

\begin{abstract} We theoretically predict that vanadium-based Janus dichalcogenide monolayers constitute an ideal platform for spin-orbit-torque memories. Using first principle calculations, we demonstrate that magnetic exchange and magnetic anisotropy energies are higher for heavier chalcogen atoms, while the broken inversion symmetry in the Janus form leads to the emergence of Rashba-like spin-orbit coupling. The spin-orbit torque efficiency is evaluated using optimized quantum transport methodology and found to be comparable to heavy nonmagnetic metals. The coexistence of magnetism and spin-orbit coupling in such materials with tunable Fermi-level opens new possibilities for monitoring magnetization dynamics in the perspective of non-volatile magnetic random access memories. 
\end{abstract}
\maketitle 
\paragraph*{Introduction - }

The search for magnetism in two-dimensional (2D) materials has experienced a complete revival in recent years, with the observation of room temperature magnetism \cite{Gong2019}. Layered van der Waals materials such as transition metal phosphorous trichalcogenides\cite{Joy1992} (MnPS$_3$, NiPS$_3$ etc.), transition metal trihalides\cite{McGuire2017} (CrCl$_3$, CrI$_3$), and transition metal dichalcogenides \cite{Freitas2015} (TMDs - VSe$_2$, CrSe$_2$, CrTe$_2$) are all known to display quasi-two dimensional magnetism and therefore have the potential for supporting magnetism down to the monolayer limit. Remarkably, 2D magnetism has been recently reported in CrI$_3$ \cite{Huang2017b}, CrSiTe$_3$ \cite{Lin2016} and Cr$_2$Ge$_2$Te$_6$ \cite{Gong2017} mono- and bilayers, as well as in Fe$_3$GeTe$_2$ \cite{Fei2018,Deng2018}, VSe$_2$ \cite{Bonilla2018}, and MnSe$_2$\cite{OHara2018} monolayers. The latter results obtained at room temperature, are particularly intriguing because TMDs are known to exhibit charge density wave instabilities, preventing the onset of magnetism \cite{Walker1983,Coelho2019}. This observation suggests that the substrate plays an important role in the stabilization of magnetism, which has yet to be confirmed experimentally. Despite this on-going debate, these discoveries open appealing perspectives for low-dimensional electronics.

In fact, among the vast zoology of 2D materials, the family of TMDs offers a versatile platform for the advancement of disruptive ultrathin electronics \cite{Xu2014,Li2019}. The advent of 2D magnetism fosters the realization of flat spintronics devices, whose operation is based on the electron's spin rather than its charge. For instance, tunneling magnetoresistance \cite{Song2018}, and spin tunnel field-effect transistors \cite{Jiang2019} have been recently realized. In addition, TMDs also have a large spin-orbit coupling \cite{Zhu2011} that has been successfully exploited for gate-controlled spin manipulation in graphene/TMD bilayers \cite{Yan2016,Dankert2017,Ghiasi2019,Benitez2020}. This feature is particularly interesting in the context of spintronics as it enables the onset of spin-orbit torque \cite{Manchon2019}. Spin-orbit torque (SOT) is a mechanism by which the orbital angular momentum is transferred to the spin angular momentum via spin-orbit coupling, thereby inducing a magnetic torque of the magnetization. It requires the presence of spin-momentum locking, which only emerges upon inversion symmetry breaking and strong spin-orbit coupling. The effect has been reported in both non-centrosymmetric magnets \cite{Chernyshov2009} and magnetic multilayers \cite{Miron2011b,Liu2012}, opening avenues for the development of non-volatile magnetic random access memories \cite{Cubukcu2018}. In this context, magnetic TMDs present an interesting paradigm for the realization of such ultrathin SOT magnetic random access memories.


In this Letter, we study the so-called Janus monolayer where the transition metal ion is embedded between dissimilar chalcogen elements. Such a structure allows for the coexistence of a finite magnetization, broken inversion symmetry and strong spin-orbit coupling, promoting a remarkable paradigm for {\it all-in-one} SOT-building block. Magnetic properties of the 2H phase of vanadium-based TMDs have been previously studied by density functional theory (DFT) \cite{Ma2012,Fuh2016,Fuh2016b,He2018}, but the more experimentally-relevant 1T phases remain to be explored in-depth. Our first principles calculations of various vanadium-based TMDs in their 1T phase unveil the impact of inversion symmetry breaking on the magnetic properties and the emergence of Rashba-like spin textures, which are key enabling features for current-driven gate-controlled SOT activation of magnetic dynamics and reversal.

\begin{figure}
\begin{center}
	\includegraphics[width=8cm]{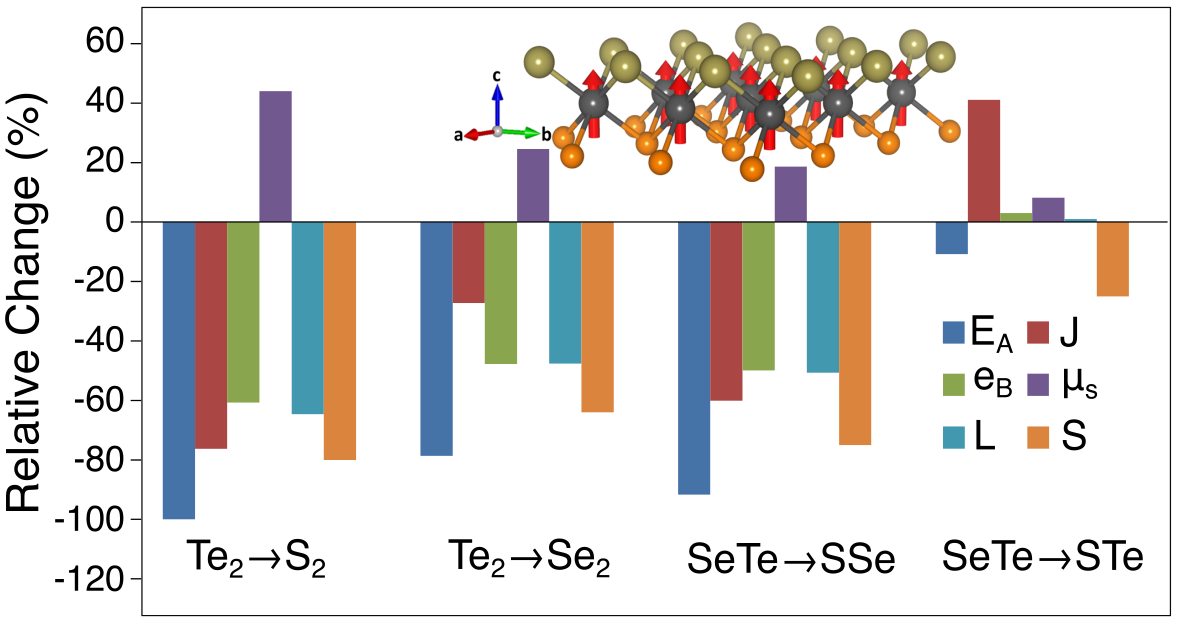}
\caption{(Color Online) Relative changes of the magnetic anisotropy $E_{\rm A}$, exchange coupling $J$, magnetic moment $\mu_{\rm s}$, Bader charge $e_{\rm B}$ and orbital angular momentum $L$ in VTe$_2$ upon substitution of Te atoms to Sulfur (S) and Selenium (Se), and also upon the transformation from the Janus form VSeTe into VSSe and VSeTe.\label{fig1} }
\end{center}
\end{figure}

\paragraph*{Method - }
We consider a VXY Janus monolayer with vanadium (V) concentrating most of the magnetic moments and octahedrally coordinated with chalcogenides ions (X and Y) in the 1T-phase, as depicted on the inset of Fig.\ref{fig1}. The DFT calculations were performed using the Perdew-Burke-Ernzerhof (PBE) method of the generalized gradient approximation (GGA) exchange-correlation functional as implemented in the Vienna Ab initio Simulation Package \cite{Kresse1993,Kresse1996a} (VASP). Grimme dispersion correction with Becke-Jonson damping (DFT-D3) has been adopted to eliminate the effect of van der Waals interactions\cite{Grimme2010,Grimme2011}. In the calculations, an energy cutoff of 600 eV is used for the plane-wave basis expansion, and total energies are properly converged with criteria of 10$^{-5}$ eV per unit cell. $\Gamma$-centered k-grids 10$\times$10$\times$1, 16$\times$16$\times$1, 24$\times$24$\times$1, and 32$\times$32$\times$1 have been sampled for the structural and magnetic calculations. A vacuum space of 15 {\AA} has been applied to avoid interaction between monolayers. The magnetic configuration has been assessed based on the force theorem by computing $E=E_{\rm AFM} -E_{\rm FM}$, where $E_{\rm AFM}$ and $E_{\rm FM}$ represent the total energies for the simple (G-type) antiferromagnetic and ferromagnetic spin arrangements, respectively. Spin texture calculations of the valence band of 1T-VXY monolayers have been investigated around the $\Gamma$ point and for the [001] orientation for model systems of 1$\times$1 boxes within three atomic layers of 1T-VXY. Crystal structures of the bulk VXY have been optimized by establishing the ground state geometries. 

The SOT efficiency is computed in the linear-response regime by using the Kubo-Bastin formula \cite{Bastin1971}
\begin{align}
\langle \hat{A}(\varepsilon) \rangle
=-2 \int_{-\infty}^\varepsilon d\varepsilon'\text{Im}\left( \text{tr}\left[ \delta(\varepsilon'-\hat{H})\hat{A}\partial_{\varepsilon'}G^+_{\varepsilon'}(\hat{\bm{J}}\cdot\bm{E}) \right]\right), \end{align}
where $\hat{A}$ is the operator driven out of equilibrium by the electric field $\bm{E}$, $\hat{\bm{J}}$ the current operator which can be expressed in terms of the Hamiltonian $\hat{H}$ and the position operator $\hat{\bm{R}}$ as $\hat{\bm{J}}\equiv -e[\hat{H},\hat{\bm{R}}]$, and $G^+_\varepsilon\equiv\lim_{\eta\rightarrow 0}1/(\hat{H}-\varepsilon+i\eta)$ is the retarded Green's function. For the spin density, we used $\hat{A}\rightarrow\hat{\bm{S}} = \frac{1}{\Omega} \sum_{i} \hat{c}_{i}^\dagger \hat{\bm\sigma}\hat{c}_{i}$ where $\hat{c}_{i}^\dagger=(c_{i,\uparrow}^\dagger,c_{i,\downarrow}^\dagger)$ and $\hat{\bm\sigma}$ is the vector of Pauli matrices for spin 1/2. The torque is computed by $\hat{A}\rightarrow\hat{\bm\tau}=\frac{1}{\Omega} {\bf m}\times\sum_{i} \Delta_i\hat{c}_{i}^\dagger \hat{\bm\sigma}\hat{c}_{i}$, $\Delta_i$ being the s-d exchange on the $i$th orbital. The Green's functions in the Kubo-Bastin formula are approximated numerically by using the Kernel Polynomial Method \cite{Garcia2015,Garcia2018, Fan2019} using 300 Chebyshev expansion moments, which is equivalent to a broadening of 54 meV for this particular system. The calculations were performed on a multiorbital tight-binding Hamiltonian by projecting the band structure on maximally-localized Wannier functions as implemented in WANNIER90 \cite{Mostofi2014}. The system was then expanded into a 40$\times$40 supercell, which amounts to 35200 orbitals in the case of VSeTe.

\paragraph*{Magnetic properties - }
We compute the magnetic properties of the full vanadium-based TMD series and found ferromagnetic order ground states for all elements. In Table \ref{tab1}, we report the magnetic properties of monolayers involving Te, which were the only one displaying a finite magnetic anisotropy energy (E$_{\rm A }$) together with the largest exchange energies ($J$) and magnetic moment ($\mu_{\rm s}$), dominated by the spin contribution ($S$). Remarkably, VTe$_2$ and VSeTe also possess the largest orbital angular momentum ($L$), which is a precursor for the formation of a Rashba-like spin-momentum locking in the Janus form. 

\begin{center}
\begin{table}
\centering
\caption{Magnetic properties extracted from fully relativistic first principle methods of selected Vanadiun-based T-VXY \label{tab1}} \
\begin{tabular}{ccccccc} \hline 
Material  &$E_{\rm A}$ (meV) & $J$ (meV)& $\mu_{\rm s}~(\mu_{\rm B})$ & $S~(\mu_{\rm B})$ & $L~(\mu_{\rm B})$\\ \hline 
VTe$_{2}$ &0.239 & 38.96 & 1.454 & 1.476 & -0.025\\ \hline 
VTeSe  &0.241 & 50.82 & 1.362 & 1.391 & -0.024 \\ \hline 
\end{tabular}
\end{table}
\end{center}
Fig. \ref{fig1} shows the effect of chalcogen substitution \cite{SuppMat}, where it is clear that lighter atoms are detrimental to the magnetic and orbital properties, suppressing the magnetic anisotropy, and reducing the exchange energy, magnetic moment and orbital momentum altogether. We found that such a behavior is associated with a depletion of the charge density on vanadium due to ionic bonding with the chalcogen elements, which is larger for sulfur chalclogen as we demonstrated by performing a Bader charge analysis ($e_{\rm B}$ in Fig. \ref{fig1}). In addition, these results suggest that vanadium-based chalcogenide monolayers have the capability for {\em easy-plane} magnetic anisotropy, which makes them an interesting platform for the realization of spin superfluidity, a remarkably efficient manner to convey spin information \cite{Takei2014}. 

\begin{figure}
 \includegraphics[width=8cm]{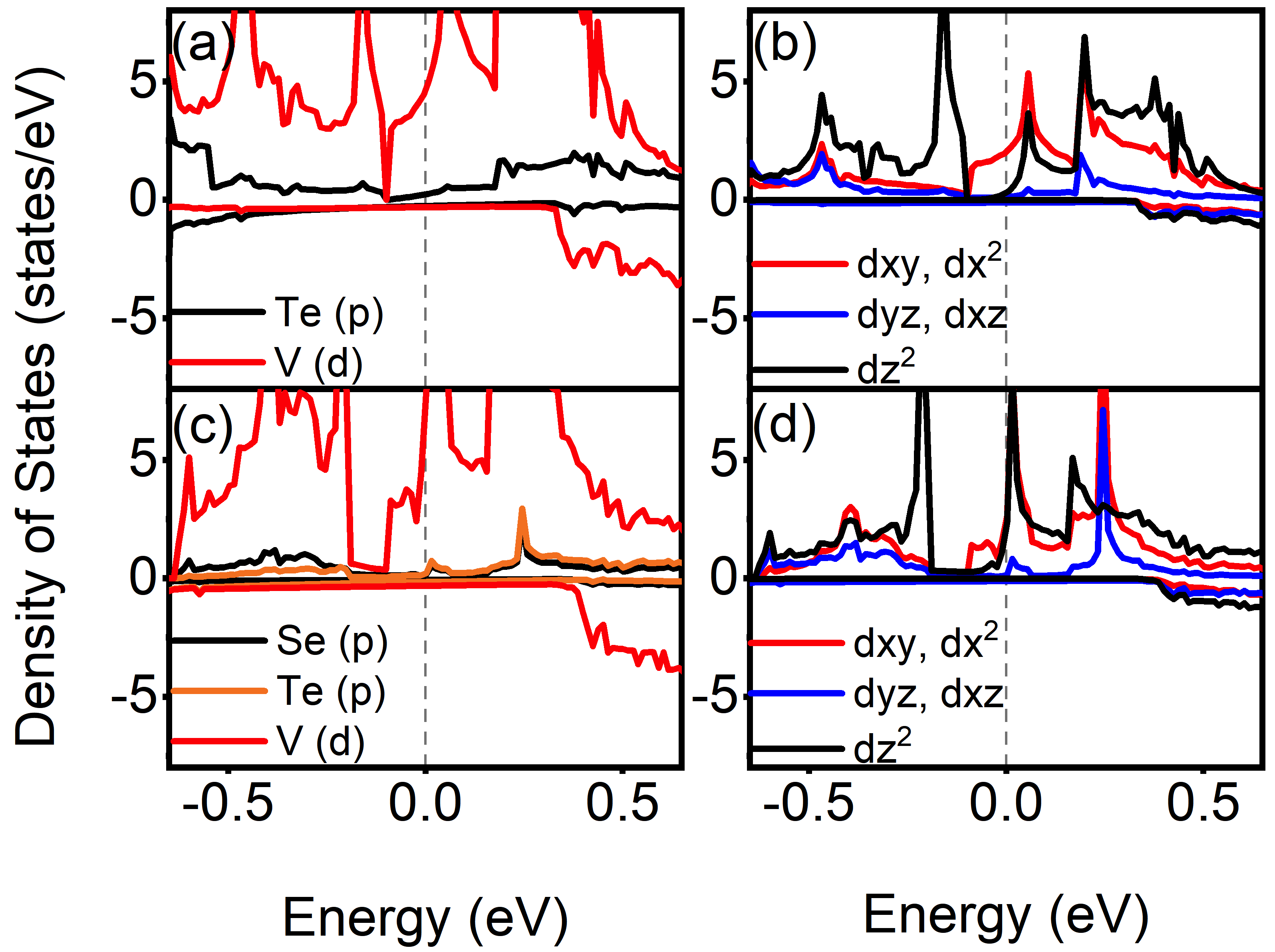}
\caption{(Color Online)Projected density of states of (a,b) VTe$_2$ and (c,d) VSeTe. The left panels display the projection of Se-p, Te-p and V-d orbitals, whereas the right panels display the orbital-resolved projection on the V-d orbitals. Positive (negative) density of states stands for up (down) spins. \label{fig3}}
\end{figure}

To identify the bonding characteristics of 1T-VXY monolayers, we compute the projected densities of states for all magnetic TMDs around the Fermi level. We determined that the contributions of the p-orbitals of the chalcogens and the d-orbitals of vanadium dominate around the Fermi level, whereas the other orbitals have negligible contributions, as seen in Fig. \ref{fig3}(a) for VTe$_2$ and Fig \ref{fig3}(b) for VSeTe. Such a result indicates that the states around the Fermi level are due to a covalent hybridization of the vanadium d-orbitals with the p-orbitals of the chalcogens. Interestingly, the density of states for down spins (negative values) is substantially reduced around Fermi energy, implying that the magnetic TMDs are in fact close to half-metallic behavior. 

We will now focus on understanding the role of the broken inversion symmetry, which reduces the point group symmetry from D$_{3d}$ to C$_{3v}$. For both groups, the d-orbitals split into three states, a (d$_{z^2}$) state, and two twofold degenerate states (d$_{zx}$, d$_{yz}$) and (d$_{x^2-y^2}$, d$_{xy}$). In Fig.\ref{fig3}(b,d), we show the density of states projected on these five d orbitals, where it is clear that (d$_{x^2-y^2}$, d$_{xy}$) is strongly suppress around the Fermi energy, while (d$_{zx}$, d$_{yz}$) are ubiquitous. Such an imbalance between the degenerated orbitals at the Fermi energy is directly associated with the emergence of a large orbital moment, and explains why Te has the largest orbital moment when compared to the other chalcogens (see Fig.\ref{fig1}). The (d$_{z^2}$) state is a consequence of p-d hybridization, hence it is strongly suppressed in VTe$_2$ [Fig. \ref{fig3}(b)] but is significant in VSeTe due to the broken inversion symmetry [Fig. \ref{fig3}(d)]. The enhanced orbital moment and p-d hybridization suggests the formation of a Rashba-like spin-momentum locking.
\par

 \par

\begin{figure}
\begin{center}
\includegraphics[width=9cm]{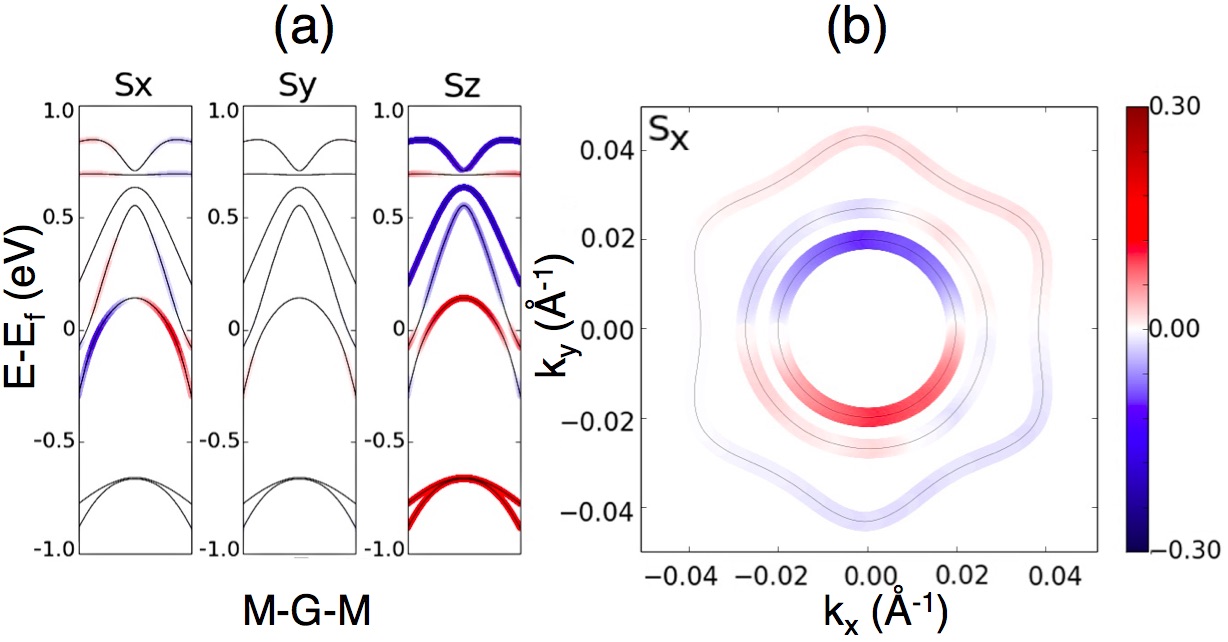}
\caption{(Color Online) (a) Spin-resolved band structure of 1T-VSeTe projected on vanadium d-orbitals. The three subpanels refer to S$_x$, S$_y$ and S$_z$ components of the spin density, the red (blue) color indicates a positive (negative) value. (b) S$_x$ component of the spin texture in momentum space projected on vanadium d-orbitals. During these calculations, the magnetization is set perpendicular to the plane. \label{fig5}}
\end{center}
\end{figure}

\paragraph*{Spin-momentum locking - }
To determine how the inversion asymmetry promotes spin-momentum locking in VSeTe, we compute its band structure along M-\ensuremath{\Gamma}-M [see Fig.\ref{fig5}(a)]. We focus on the two bands crossing the Fermi level with hole-like dispersion. These bands present simultaneous signatures of spin-momentum locking and exchange splitting. After analyzing the orbital nature of the bands \cite{SuppMat}, we determined that vanadium d-orbitals are present across the whole band structure and lead to strong spin-polarization driven by near half-metallicity, which also coexists with a Rashba-like spin-momentum locking antisymmetric in $k$, as evidenced in Fig. \ref{fig5}(b). The simultaneous presence of magnetic moments and Rashba like interactions enables current-driven SOTs, and we found that VSeTe has the best potential for strong SOT owing to its largest odd-in-momentum spin texture \cite{SuppMat}. Since, the band-states are found to be mainly dominated by hybridized d$_{z^2}$ orbital from vanadium and $p_z$ orbitals from selenium, we further developed a simpler tight-binding model, evidencing that the Rashba-like interaction stems from the admixture of d$_{z^2}$ and (d$_{zx}$, d$_{yz}$) orbitals. At second order in perturbation theory, the d$_{z^2}$ orbital reads \cite{SuppMat}
\begin{eqnarray}
|d_{z^2}\rangle_p&=&\left(1-\frac{1}{2}\sum_\alpha\frac{|V_{z^2,z}^{\alpha}|^2}{(\varepsilon_{d_{z^2}}-\varepsilon_\alpha)^2}\right)|d_{z^2}\rangle_0\\
&&+\sum_{s,\alpha}\frac{V_{s,z}^{\alpha}V_{z,z^2}^{\alpha}}{(\varepsilon_{d_{z^2}}-\varepsilon_s)(\varepsilon_{d_{z^2}}-\varepsilon_\alpha)}|s\rangle_0\nonumber.
\end{eqnarray}
where, $V_{z^2,z}^\alpha$ ($V_{s,z}^\alpha$) the hopping integral between d$_{z^2}$ ($s$) and p$_z$ orbital from chalcogen $\alpha$ and $\varepsilon_\eta$ is the energy of orbital $\eta$. We computed model's parameters in the two-center Slater-Koster approximation. Close to the $\Gamma$-point, the spin-orbit coupling energy ${\cal H}_{\rm so}$ of this perturbed state is given by \begin{eqnarray}
\langle d_{z^2}|{\bf L}\cdot{\bf S}|d_{z^2}\rangle_p\approx \alpha_{\rm R}\hat{\bm\sigma}\cdot({\bf z}\times{\bf k})
\end{eqnarray}
where $\alpha_{\rm R}$ the Rashba parameter, which can be expressed in terms of the Slater-Koster parameters \cite{SuppMat}. This expression demonstrates that coupling the d-orbitals of vanadium with dissymmetric chalcogen bonding induces Rashba spin-orbit coupling, in agreement with the first principles calculations discussed above. The large spin splitting implies that only one spin chirality contributes to the spin transport, making this systems optimal for SOT. 

\paragraph*{Spin-orbit torque - }
After establishing the formation of Rashba-like spin-texture coupled to the magnetic moments in the Janus structure, we now evaluate its impact on the SOT efficiency. By considering the C$_{3v}$ symmetry of the Janus monolayer and applying the Neumann's principle to the current-driven field tensor \cite{Zelezny2017}, one determines that the torque possesses two components,
\begin{eqnarray}\label{sotsym1}
H_{\rm FL}&=&\chi_{\rm FL}{\bf z}\times{\bf E}\\\label{sotsym2}
H_{\rm DL}&=&\chi_{\rm DL}^1{\bf m}\times({\bf z}\times{\bf E})+\chi_{\rm DL}^2m_z{\bf E}\\
&&+\chi_{\rm DL}^2[(m_yE_x+m_xE_y){\bf x}+(m_xE_x-m_yE_y){\bf y}].\nonumber
\end{eqnarray}
The first terms ($\chi_{\rm FL},\chi_{\rm DL}^1$) are similar to the one expected from the standard theory of the Rashba gas, while the third damping-like term ($\chi_{\rm DL}^2$) arises from the threefold rotation combined with the mirror symmetry. A similar term was obtained by \citet{Johansen2019} for Fe$_3$GeTe$_2$, and one can show that it derives from a potential $E_{\rm T}=\chi_{\rm DL}^2(m_xm_yE_x+(m_x^2-m_y^2)E_y/2)$, which therefore expresses a current-driven magnetic anisotropy \cite{Johansen2019}. Importantly, in contrast to Fe$_3$GeTe$_2$, the obtained torque in 1T-VSeT coexists with the conventional "Rashba" torques and therefore has the ability to modify the overall magnetization dynamics and even permits {\em zero-field switching}. 

We note that in a realistic system, this zero-field switching torque arises from Bloch states experiencing the $C_{3v}$ crystal field, so for states typically lying {\em away} from $\Gamma$-point. In vanadium chalcogenides, the magnetic and transport properties are dominated by the $\Gamma$-point physics [Fig. \ref{fig5}(b)], which suggests that these properties shall display cylindrical symmetry at the lowest order, and that high-order effects, such as the derived current-driven magnetic anisotropy or the threefold magnetocrystalline anisotropy expected in hexagonal magnets, should only emerge as a small correction. As a matter of fact, our calculation of the magnetic anisotropy cannot resolve any threefold planar component, so that the value of $\chi_{\rm DL}^2$ should be vanishingly small.

\begin{figure}
\begin{center}
\includegraphics[width=8cm]{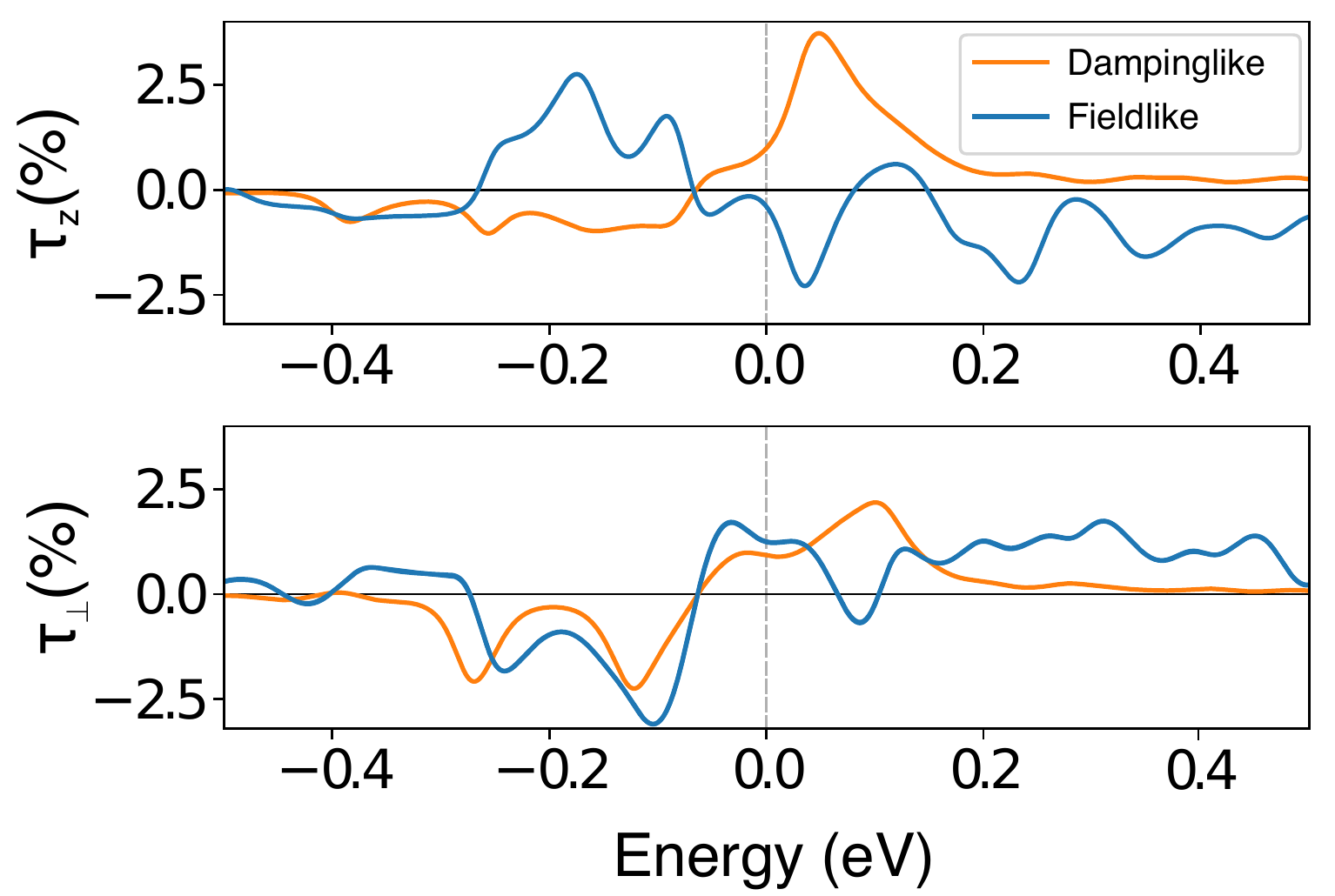}
\caption{(Color Online) Current-driven torque efficiency as a function of the transport energy, when the magnetization lies perpendicular to the plane (top) and in plane (bottom). The figure shows both the dampinglike (orange) and fieldlike (blue) components. \label{sot}}
\end{center}
\end{figure}

Figure \ref{sot} shows the energy dependence of the torque efficiency when the magnetization points out of plane ($\tau_z$ - top) and in the plane ($\tau_\bot$ - bottom), computed by setting the electric field along ${\bf x}$. We verified that similar results are obtained when setting the electric field along ${\bf y}$. The torque efficiency is defined as the torque per unit electric field divided by the longitudinal conductivity, $\chi_{\rm FL,DL}/\sigma_{c}$, and is unitless. We find that the torque adopts the conventional Rashba form, Eqs. \eqref{sotsym1}-\eqref{sotsym2}, displaying both dampinglike (orange) and fieldlike torques (blue) with a strong anisotropy when changing the magnetization direction from in-plane to out-of-plane. Additional components with one order of magnitude smaller than the conventional torques are observed and discarded. The overall magnitude of the torque ($\tau_{z,\bot}<2.5~\%$) remains limited compared to the best transition metal substrates such as Pt ($\sim8~\%$) and W ($\sim30~\%$) (see Ref. \cite{Manchon2019}). However, it is sufficient to switch the magnetization direction. We further determined that the current density required to overcome VSeTe anisotropy is 
 $J_c^0\approx 7.5\times 10^7$ A/cm$^2$ \cite{SuppMat}. This critical current density is surprisingly small in spite of the weak spin-orbit coupling of vanadium (in fact, one of the weakest in the transition metal series). Besides, $J_c^0$ has been computed at zero temperature and will be naturally further reduced by thermal activation in experimental conditions. Another remarkable feature is the possibility to {\em quench} the magnitude of the torque by tuning the Fermi level. Indeed, Fig. \ref{sot} shows that $\tau_z$ ($\tau_\bot$) is maximum around 50 meV (-100 meV) from Fermi level. As a result, by tuning the Fermi level either by doping or using the more technologically relevant ionic gate voltage control, one can substantially enhance the efficiency of the torque and turn it on and off at will, thereby implementing a {\it gate-controlled spin switch}. Interestingly the magnetic properties of 2D materials are usually particularly sensitive to surface engineering and substrate effects, enabling easier tuning of the magnetic anisotropy \cite{Park2020,Kim2020}. 

\paragraph*{Conclusion - }
 We studied monolayers vanadium-bases magnetic TMDs in their Janus form as {\it all-in-one} SOT building block, allowing for electrically-controlled magnetism in a single material. Among the different candidates, we found that VSeTe is the most promising due to possessing the largest magnetic anisotropy, exchange interaction, and Rashba-like behavior. We evaluated how these features translate into the magnetization dynamics by computing the current-driven torque using Kubo formula and demonstrated that it is large enough to realize switching at room temperature. Moreover, we also show that such a switching can be tuned by electronic or ionic gating, opening fascinating perspectives for applications such as SOT magnetic memories and spin-charge conversion devices. 

\acknowledgments
The authors were supported by King Abdullah University of Science and Technology (KAUST) through the award OSR-2018-CRG7-3717 from the Office of Sponsored Research (OSR). This work used the resources of the Supercomputing Laboratory at King Abdullah University of Science and Technology (KAUST) in Thuwal, Saudi Arabia. ICN2 authors were supported by the European Union Horizon 2020 research and innovation programme under Grant Agreement No. 785219 (Graphene Flagship), by the CERCA Programme/Generalitat de Catalunya, and by the Severo Ochoa program from Spanish MINECO (Grant No. SEV-2017-0706 and MAT2016-75952-R).  
\bibliography{Biblio-Idris}

\end{document}